\documentclass[10pt,letterpaper,twocolumn,english,prb,showpacs]{revtex4-1}
\usepackage[T1]{fontenc}
\usepackage[latin9]{inputenc}
\usepackage{amsmath}
\usepackage{amssymb}
\usepackage{graphicx}
\usepackage{babel}

\pdfoutput=1

\begin{document}

\title{Localized quasiholes and the Majorana fermion in fractional quantum
Hall state at $\nu=5/2$ via direct diagonalization}

\author{M. Storni and R. H. Morf}

\affiliation{Condensed Matter Theory, Paul Scherrer Institute, CH-5232 Villigen,
Switzerland}

\date{\today}

\begin{abstract}
Using exact diagonalization in the spherical geometry, we investigate
systems of localized quasiholes at $\nu=5/2$ for interactions interpolating
between the pure Coulomb and the three-body interaction for which the
Moore-Read state is the exact ground state. We show that the charge
$e/4$ quasihole can be easily localized by means of a $\delta$-function
pinning potential. Using a tuned smooth pinning potential, the quasihole
radius can be limited to approximately three magnetic length units. For systems
of two quasiholes, adiabatic continuity between the Moore-Read and
the Coulomb limit holds for the ground state, while for four quasiholes,
the lowest two energy states exhibit adiabatic continuity. This implies
the existence of a Majorana fermion for pure Coulomb interaction.
We also present preliminary results in the Coulomb limit for braiding
in systems containing four quasiholes, with up to 14 electrons, diagonalizing
in the full spin-polarized sector of the second Landau-level Hilbert
space. 
\end{abstract}


\maketitle

\section{Introduction}

The possibility of realizing non-Abelian braiding statistics \cite{banff}
in a condensed-matter system has generated a great deal of interest
both for theoretical and experimental studies. One such system that
may be closest to experimental realization is the fractional quantum
Hall (FQH) state at filling fraction $\nu=5/2$, first observed by
Willett \textit{et al.} \cite{Willett87} Whether the experimentally
observed $5/2$ state indeed has charged excitations with non-Abelian
braiding statistics is still unknown, altough the discovery of a neutral
current in experiment may make us hopeful.\cite{bid10} For a review,
we refer the reader to Nayak \textit{et al.} \cite{NayakRMP08} and
to Stern.\cite{stern10}

In an earlier work \cite{prl} we presented numerical evidence that
the FQH ground state (GS) at $\nu=5/2$ for Coulomb interaction, when
spin polarized, is in the same universality class as the non-Abelian
Moore-Read (MR) Pfaffian state.\cite{MooreRead91} Calculating the
energy spectrum of the few lowest-energy states by exact diagonalization
for electron interactions interpolating between the Coulomb interaction
$V_{C}$ and the three-body interaction $V_{3b}$, for which the MR
state is the unique GS, we found for all examined system sizes adiabatic
continuity (AC) of the GS and no sign of a decrease of the gap between
the two limits $V_{C}$ and $V_{3b}$. We concluded that AC can be
expected even in the thermodynamic limit. In addition, we drew a phase
diagram in the two-body interaction space (in the vicinity of the
Coulomb interaction), showing that the FQH gapped phase coincides
with the MR phase.

Previous theoretical work \cite{Wen93,Wan,moller} gave indications
for AC for the spin-polarized $\nu=5/2$ GS between the MR and the Coulomb
limit. In the disk geometry Xin Wan \textit{et al.} \cite{Wan} were
also able to localize a single quasihole (QH) at the center of the disk
showing
that, in a certain range of QH pinning and system confinement potential
strength, the lowest-lying state belongs to the total angular momentum
value that corresponds to a MR $e/4$ charged QH, for electron interactions
interpolating between $V_{3b}$ and $V_{C}$.

In spite of the AC of the GS, the question whether the elementary
charged excitations preserve their non-Abelian properties in going
from the MR to the Coulomb limit still needs to be examined: This
is a very important issue, because braiding of non-Abelian quasiparticles
has been proposed for topological quantum computation.\cite{DasSarma05,NayakRMP08}
At $\nu=5/2$, one expects that QHs can have either
charge $e/4$ with the possibility of having non-Abelian braiding
statistics or $e/2$ with Abelian fractional statistics. Indeed, the
results by T\H{o}ke \textit{et al.} \cite{TokeQH} have cast doubt
on the existence of localized QHs with charge $e/4$.

Here, we study fully spin-polarized systems with localized QHs in
the spherical geometry. This polarization choice is motivated by theoretical
investigations concerning this issue: One of us showed that the GS
of the disorder-free FQH state at $\nu=5/2$ is spin polarized even
for vanishing Zeeman energy;\cite{Morf98} recent theoretical work
\cite{dimov08,feiguin09} confirmed this result. However, from the
experimental point of view, the situation is less clear. Transport
experiments with variable electron density \cite{pan,nuebler} or
in a tilted magnetic field at high density \cite{chizhang10} have
been interpreted in controversial ways: Varying the electron density
by more than a factor $2$,\cite{pan} the excitation gap varies smoothly
with no break in slope or discontinuity, indicating that neither GS
nor excitations change their character while the Zeeman energy varies
by an amount that exceeds the measured energy gap by a factor of $5-10$.
That an unpolarized state would survive under this condition
appears unlikely. Also, in high-density samples, the disappearence
of the gapped phase in a tilted magnetic field cannot be explained
by a Zeeman energy term with a GaAs bulk $g$ factor.\cite{chizhang10}
Nevertheless, arguments have been presented that these experimental
results might be consistent with a spin-unpolarized $5/2$ state.\cite{dassarma10}
Also first results from direct spin polarization measurements, using
optical methods,\cite{mstern10,rhone10} appear to be consistent
with a spin-unpolarized state. As the experimental samples are likely
to be inhomogeneous and consist of compressible regions separated
by percolating incompressible filaments at $\nu=5/2$,\cite{ambrumenil10}
it is not clear that the optical data relate solely to the filaments
or to the compressible regions.\cite{rhone10} The separate determination
of the polarization state of the incompressible parts has not yet
been possible.
In addition, charged impurities may also induce skyrmions in the incompressible
domains,\cite{wojs10,rhone10} which will decrease the total spin
polarization. We also note that, once thermally assisted tunneling
is taken into account,\cite{ambrumenil10} the gap obtained from
analyzing the dissipative conductance can be as consistent with expectations
for a spin-polarized system at $\nu=5/2$ (Refs. 17, 24 and 25)
as for the hierarchy state at $\nu=4/9$.

In our investigations at first we show that charge $e/4$ QHs can
be localized by a $\delta$-function potential and, varying the form
of the localization potential away from the simple $\delta$-function,
we also show that there is an optimal localizing potential that minimizes
the density oscillations around the QHs. We also find that in systems
with two QHs, in the Coulomb case, it is sufficient to have a single
$\delta$-function: This localizes a single QH in the GS, while the
other is automatically at the antipodal point on the sphere, because
of the Coulomb repulsion.

We then study how states containing two and four QHs at $5/2$ evolve
as the interaction is varied between $V_{3b}$ and $V_{C}$, and show
that there is AC between the two interaction limits, as in the previous
work.\cite{prl}

For two QHs, as the interaction is varied from $V_{3b}$ to $V_{C}$,
we find that the lowest energy state does not mix with the higher
lying ones. For systems containing four QHs, the two lowest-lying
states forming a degenerate doublet in the $V_{3b}$ limit \cite{nayak96,bonderson11}
remain the two lowest-energy states even in the limit of pure Coulomb
interaction $V_{C}$. Thus, the Majorana fermion associated with the
doublet \cite{read00} appears to survive in the Coulomb limit.

To better study the MR limit we then repeat the investigations for
four QHs using a different pinning potential: Also with this approach
we obtain AC for the lowest-lying doublet between the two interaction
limits, confirming the above results. Finally we present some preliminary
results of braiding with QHs, showing that exchanging the position
of two QHs, in the presence of two others fixed QHs, the system
goes from one of the MR doublet states to the other, showing non-Abelian
braiding statistics. Furthermore, we investigate the fusion of two
QHs, getting an estimate for the MR doublet splitting.\cite{baraban2}

\begin{figure*}[t!]
 \includegraphics[width=1\textwidth]{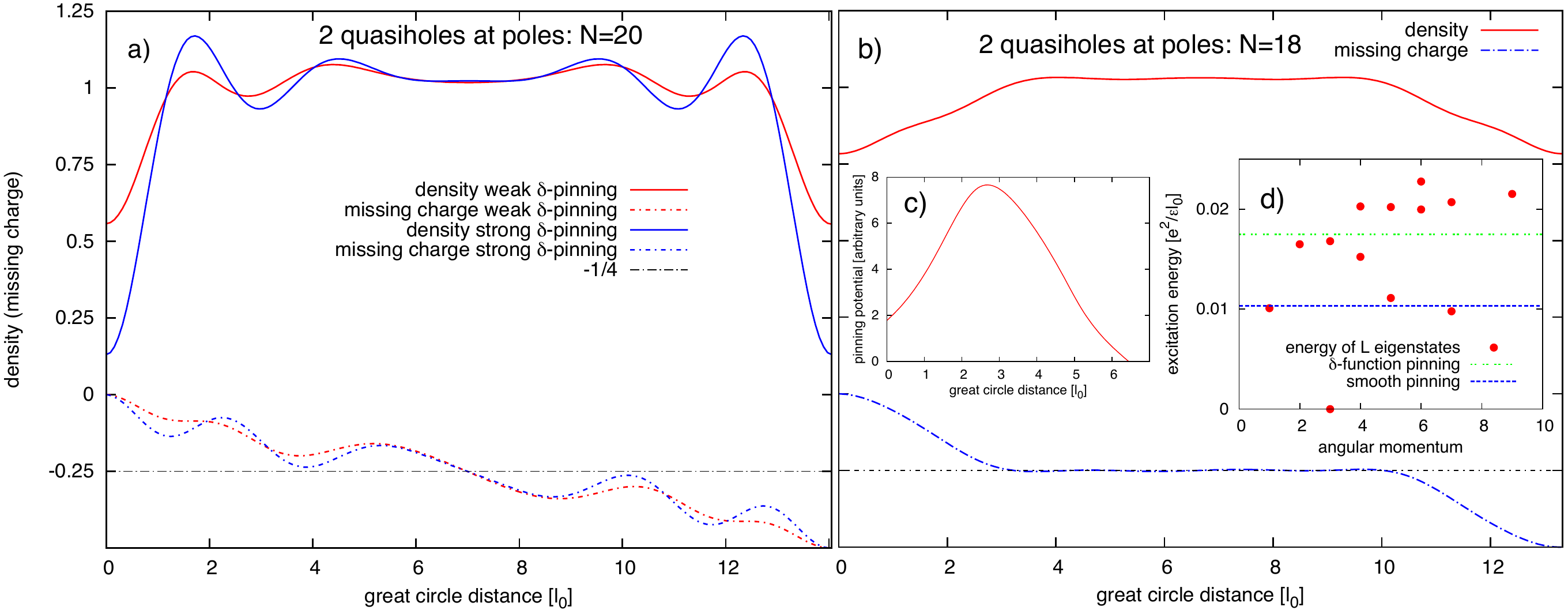}

\caption{(Color online) Polar angle dependence of electron density and missing
charge for $\nu=5/2$ systems with two QHs localized at the poles
of the Haldane sphere, for Coulomb interaction. In (a) the QHs are
localized by a $\delta$-function potential $\mathcal{V}_{pin}$ [setting
$x=0$ in Eq.~(\ref{Vpin1})], for two different pinning strengths;
in (b) by a smooth pinning potential, whose radial dependence is shown
in (c). (d) Energy spectrum of a system with two QHs, without pinning,
as function of the total angular momentum $L$, and comparison with
the Coulomb energy for $\delta$ function and smooth pinning [cf.~the
text for more details; results are for $N=20,\, N_{\phi}=38$
in (a) and for $N=18,\, N_{\phi}=34$ in (b),(c) and (d)]. }
\end{figure*}

\section{The system}
\label{system}

In the following we consider fully spin-polarized electron systems
on the surface of a sphere:\cite{Haldane} By exact diagonalization
we obtain their low-lying states and energy spectra. In the spherical
geometry for a half filled Landau level (LL) the number of electrons
$N$ and the number of flux quanta $N_{\phi}$ are related by $N_{\phi}=2N-S$,
where the shift $S$ (Ref. 31)
is a topological quantum number
that depends on the particular FQH state.

For our investigations of the FQH at filling factor $\nu=5/2$ we
study a half filled second LL, while the two lower-lying filled levels
are considered as inert. For such a system the GS is at shift $S=3$
(as obtained by exact diagonalizations \cite{Morf98}), the same value
as for the MR state.\cite{greiter} We can insert $N_{QH}$ $e/4$-charged
QHs into the system by adding $\frac{1}{2}N_{QH}$ supplementary
flux quanta, i.e., 
\begin{equation}
N_{\phi}(N,N_{QH})=2N-3+N_{QH}/2.
\end{equation}

\begin{figure}[t!]
 \includegraphics[width=1\columnwidth]{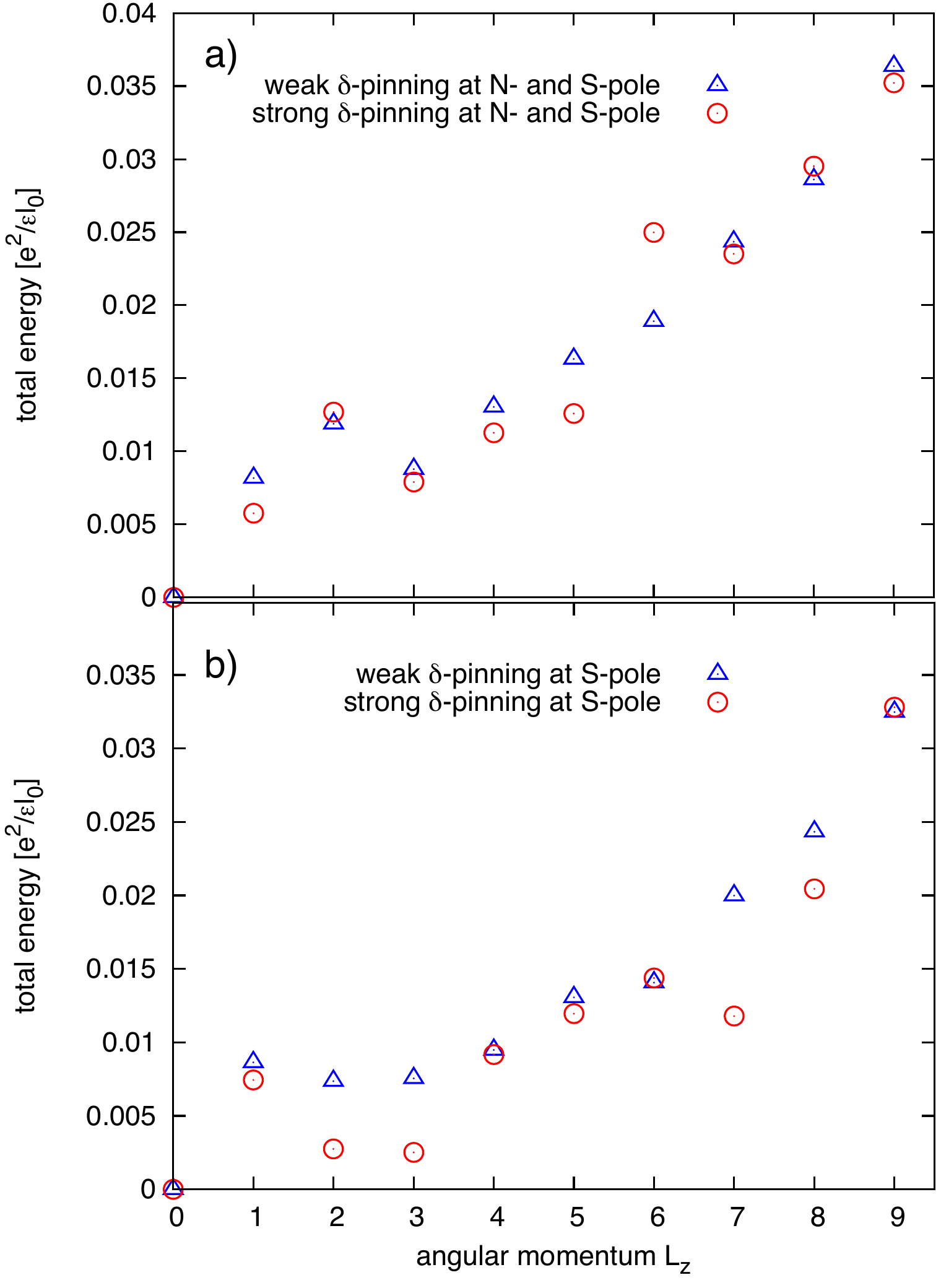}

\caption{(Color online) (a) Energy spectrum of a two-QH system as a function
of the angular momentum component $L_{z}$ when $\delta$ functions
are inserted at both poles (for two different pinning strengths, $N=16,\, N_{\phi}=30$);
the lowest-energy state is the reference. (b) The same for a single
$\delta$ function at the south pole.}
\end{figure}

In the spirit of our previous work \cite{prl} we consider interactions
of the form 
\begin{equation}
V=(1-x)V_{C}^{(1)}+xV_{3b}^{(0)}+V_{pin}(x),\label{eq:1}
\end{equation}
where $0\le x<1$, interpolating between the Coulomb interaction
$V_{C}$ and the three-body interaction $V_{3b}$, whose exact GS
is the MR wave function,\cite{Wen93} 
\begin{equation}
V_{3b}=\frac{A}{N^{5}}\sum_{i<j<k}^{N}S_{ijk}\left\{
\Delta_{j}\delta(i-j)\Delta_{k}^{2}\delta(i-k)\right\} ,\label{eq:2}
\end{equation}
when projected onto the lowest (spin-polarized) LL. Here $S_{ijk}$
expresses the full symmetrization over the permutations within the
triplet $(ijk)$ and $\delta(i-j)$ is the $\delta$-function in the
separation of particles $i$ and $j$. The constant $A$ is chosen
as in the earlier work,\cite{prl} such that the gap for $V_{3b}$
is approximately the same as that for $V_{C}$ (note that $V_{3b}$ induces
an extensive energy for all GS at any $\nu>1/2$).

The superscripts $(0)$ and $(1)$ in Eq.~(\ref{eq:1}) indicate
that the Coulomb energy $V_{C}^{(1)}$ is evaluated in the $n=1$
LL, as required for the experimentally realized $\nu=5/2$ state,
while the three-body interaction $V_{3b}^{(0)}$ is evaluated in the
$n=0$ LL, since the MR state is the GS of $V_{3b}$ in the lowest
LL.

The pinning term $V_{pin}(x)$ localizes the QHs at fixed positions
on the surface of the sphere. In our first calculations (in sections
\ref{loc} and \ref{AC}) we consider a parametrization of the pinning
potential of the form 
\begin{equation}
V_{pin}(x)=(1-x)\cdot\bigl[(1-x)\mathcal{V}_{pin}^{(1)}+x\mathcal{V}_{pin}^{(0)}\bigr],\label{Vpin1}
\end{equation}
where the pinning function $\mathcal{V}_{pin}$ is still to be chosen,
such that it localizes exactly an $e/4$-charged QH. This $V_{pin}(x)$
couples to the electron density, interpolating like the electron interaction
between LL $n=1$ and $n=0$ with weights $(1-x)$ and $x$, respectively.
The overall factor $(1-x)$ makes the localization term vanish in
the MR limit as $x\to1$, just like the Coulomb term, both helping to
separate the energies of states with charge $e/4$ QHs.

However, in the MR limit ($x\rightarrow1$), the $e/2$-charged ``double
QH'' has a zero in the electron density and thus zero pinning energy,
while the $e/4$-charged QH shows a finite local minimum of the density
and thus finite pinning energy. To separate $e/4$-charged QHs requires
either an additional repulsive interaction between QHs or a special
pinning potential coupling to the monopole and quadrupole moment of
the QHs (cf. problems discussed in T\H{o}ke \textit{et al.} \cite{TokeQH}).
To avoid this difficulty, in this first series of calculations we
never reach exactly the MR point, but approach it closely enough to
identify the lowest-energy states. In Section \ref{MRlimit} we will
then use a different localization potential \cite{locpot} to circumvent
this problem and study the full AC up to the MR point.

All energies are measured in the usual units $e^{2}/\varepsilon\ell_{0}$,
where $\ell_{0}=\sqrt{\hbar c/eB}$ is the magnetic length, and taking
the energy of the lowest state as reference.

\section{Localization of quasiholes}
\label{loc}

We start by studying the QH localization in the pure Coulomb limit,
taking for $\mathcal{V}_{pin}$ a $\delta$-function pinning potential:
Choosing a suitable pinning strength we see that it is actually possible
to obtain separated $e/4$-charged QHs, in contrast to the results
of T\H{o}ke \textit{et al.}\cite{TokeQH} Figure 1(a) shows the
polar angle dependence of the density (upper curves) and of the integrated
missing charge (lower curves) for two different pinning strengths,
for a system containing two QHs with localizing $\delta$ functions,
one at the north and the other at the south pole of the sphere. We
observe that the two QHs are localized at the poles. However they
are large in extent and are accompanied by significant and slowly decaying
density oscillations (cf. Nuebler \textit{et al.} \cite{nuebler}).
The missing charge $e/4$ is reached (for the stronger pinning potential)
at $R\approx4\ell_{0}$ and at the equator as required by symmetry.

As the system is compressible at the position of a QH, the shape of
the pinning potential will influence its structure. We therefore study
how the pinning potential can be tuned to minimize the size of the
QH and to suppress the density oscillations around it. With the choice
of shape, depicted in Fig.~1(c), we obtain an electron density without
oscillations, as illustrated in Fig.~1(b): The smooth potential allows
to properly separate the two QHs and to reduce their radius to a
minimum of $\sim 3\ell_{0}$.

How is this possible? Figure 1(d) shows a comparison of the interaction
energies, not including the energy due to the pinning forces, for
$\delta$-function pinning (upper line) and for smooth pinning (lower
line). The circles depict the spectrum of a system with two free QHs,
i.e., without pinning force, as function of the total angular momentum
$L$. We see that the use of a smooth pinning potential allows to
obtain a smaller interaction energy, admixing only the lowest-lying
$L$ eigenstates, while the $\delta$-function potential admixes also
higher-lying states, giving rise to the large density oscillations.
We also note that a ``better'' localization, with a pinning potential
that further lowers the interaction energy, is unlikely to be possible,
because a minimal number of low-energy $L$ eigenstates is needed
to obtain the desired QH localization.

We note that the energy of two-QHs systems is affected by how well
the density oscillations fit with the separation distance of the QHs,
such that the energy will show oscillations as function of the QH separation.
The energy oscillations observed in the variational calculation of
the Majorana fermion energy \cite{baraban2} may to some extent be
caused by such commensuration effects. An optimized localization potential
might help to reduce them, hopefully allowing a more accurate calculation
of the coherence length associated with the Majorana fermion.

\begin{figure}[t!]
 \includegraphics[width=1\columnwidth]{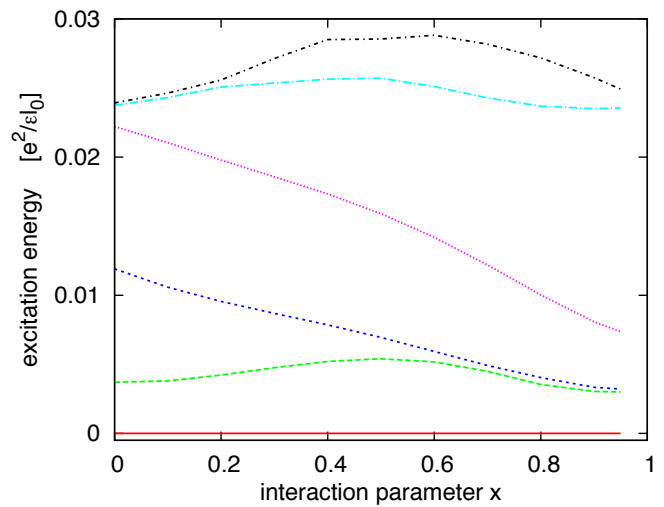}

\caption{(Color online) Low-lying energy spectrum as function of the interaction
parameter $x$ for a system of $N=12$ electrons with two QHs localized
at the poles by $\delta$-function pinning potentials; the lowest-energy
state is the reference.}
\end{figure}

\begin{figure*}[t!]
 \includegraphics[width=1\textwidth]{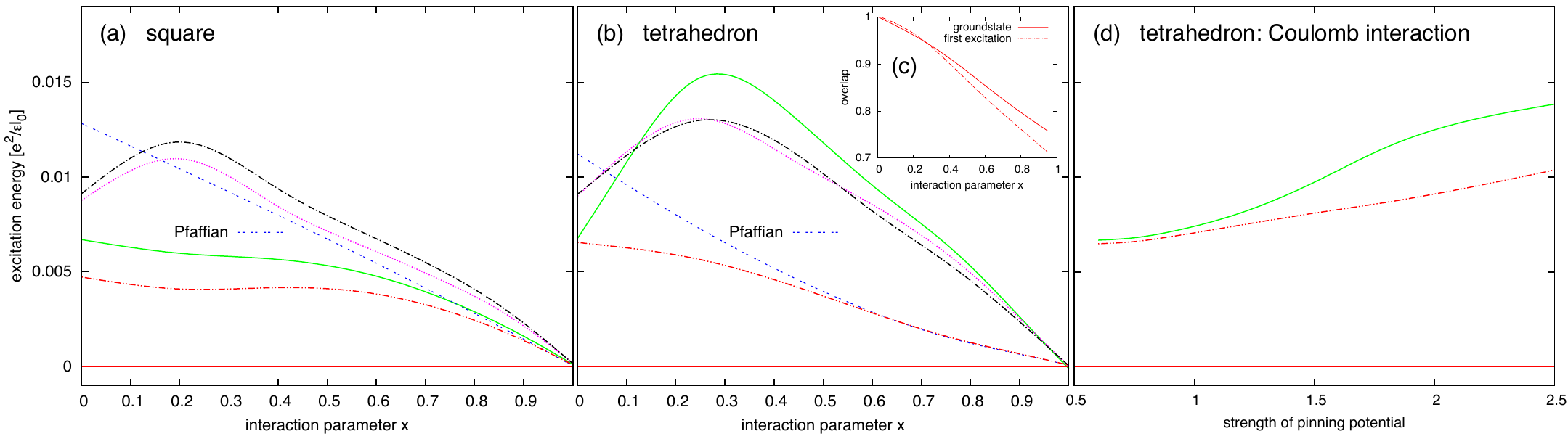}

\caption{(Color online) Low-lying energy spectrum as function of the interaction
parameter $x$ for systems with four QHs localized with $\delta$-function
pinning potentials: (a) At the corners of a square on a great circle
and (b) at the vertices of a tetrahedron; the curves labeled Pfaffian
give the variational estimate of the doublet splitting obtained using
the GS doublet for the MR limit. (c) Overlaps of the two doublet-states
for Coulomb interaction with the corresponding states for interaction
parameter $x$. (d) Low-lying energy spectrum for Coulomb interaction,
as function of the pinning strength, for four QHs in tetrahedral
position. (The lowest-energy state is the reference; $N=14$, $N_{\phi}=27$.)}
\end{figure*}

In the following, we use $\delta$-function pinning potentials. We
first study what happens if only a single $\delta$-function is inserted
in a system of $N=16$ electrons with two QHs, and whether they are
still separated or if a single $e/2$-charged ``double quasihole''
is formed. Figure 2(a) shows the energy spectrum as function of the
angular momentum component $L_{z}$ when two $\delta$ functions are
inserted at the poles, for weak and strong pinning - cf.~Fig.~1(a).
In both cases the lowest-energy state has $L_{z}=0$ and describes
two separated QHs localized at the poles. Figure 2(b) shows the same,
but for a single $\delta$ function at the south pole. The lowest
energy state has also $L_{z}=0$, indicating that the QHs are still
at maximal separation, which minimizes their Coulomb repulsion: The
QH pinned at the south pole repels the other to the north pole of
the sphere. For increasing pinning strength some states with $L_{z}\ne0$
are reduced in energy. With strong enough pinning, a state with $L_{z}\ne0$,
in which QHs would cease to be maximally separated, may become the
lowest energy state. Of particular interest is the state at $L_{z}=N/2=8$
which corresponds to a ``double QH'' with charge $e/2$ localized
at the south pole, but no charge deficiency at the north pole. At
$\nu=5/2$, for Coulomb interaction, this ``double QH'' state
has quite large interaction energy and, except perhaps in the limit
of very strong pinning, is unlikely to become the GS.

\section{Adiabatic continuity}

\subsection{Quasiholes localization with $\delta$-function pinning}
\label{AC}

We now turn to the study of AC for systems containing localized QHs.
We begin with two QHs, pinned at the poles by $\delta$ functions,
and we vary the particle interaction as in Eq.~(\ref{eq:1})
from $x=0$ (the Coulomb case) to near $x=1$ (the MR limit). Figure
3 shows the evolution of the excitation energies as function of the
parameter $x$ for $N=12$ electrons and $N_{\phi}=22$ flux units;
other system sizes give similar results. The horizontal line at zero
energy represents the GS energy: Clearly none of the lines for the
excited states come close to it and there is no mixing between them.
Thus there is no phase transition in going from the Coulomb case to
the MR limit, exactly as observed for systems without QHs.\cite{prl}

Next we investigate a system of $N=14$ electrons and $N_{\phi}=27$
flux quanta that contains $N_{QH}=4$ localized QHs.\cite{maximum_size}
This is particularly interesting because in the MR limit the GS is
a degenerate doublet,\cite{gsdegeneracy} which is associated with
a Majorana fermion \cite{read00} and with the non-Abelian braiding
statistics of the QHs:\cite{nayak96,bonderson11} For any four-QH configuration
there are two linearly independent wave functions that describe it;
braiding QHs around each other induces a linear transformation in
the degenerate subspace of the doublet. Does this doublet survive
in the Coulomb limit?

Figures 4(a) and 4(b) show results for four QHs localized at the
corners of a square on a great circle and at the vertices of a tetrahedron,
respectively. In both cases we see that the two lowest-energy states
close to the MR limit, i.e., the GS at energy $0$ and the first excited
state, remain the lowest-energy states over the entire parameter interval,
even at the Coulomb point $x=0$. During the whole evolution there
is neither crossing nor mixing with higher-lying states, indicating
the absence of a phase transition. Note that in the MR case, at $x=1$,
all the considered states have zero energy, because the pinning potential
$V_{pin}(x)$ vanishes at this point.

\begin{figure}[b!]
 \includegraphics[width=8.3cm]{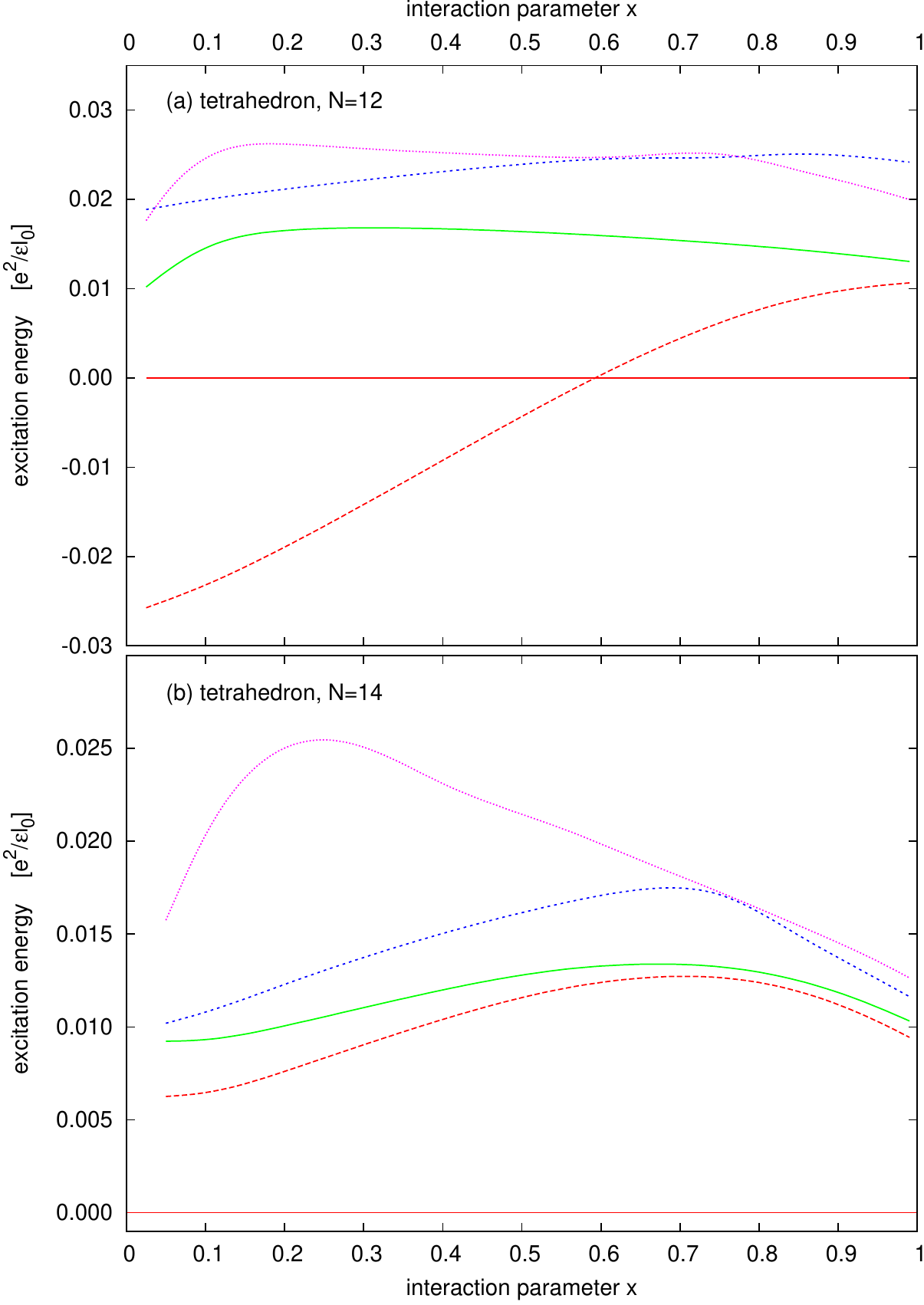}

\caption{(Color online) Low-lying energy spectrum as function of the interaction
parameter $x$ for systems with four QHs localized at the vertices
of a tetrahedron using the pinning potential in Eq.~(\ref{e/4}):
(a) $N=12,N_{\phi}=23,q=-e/4$, (b) $N=14,N_{\phi}=27,q=-0.19e$ (the
lowest-energy state in the MR limit is the reference).}
\end{figure}

\begin{figure*}[t!]
 \includegraphics[width=15.8cm]{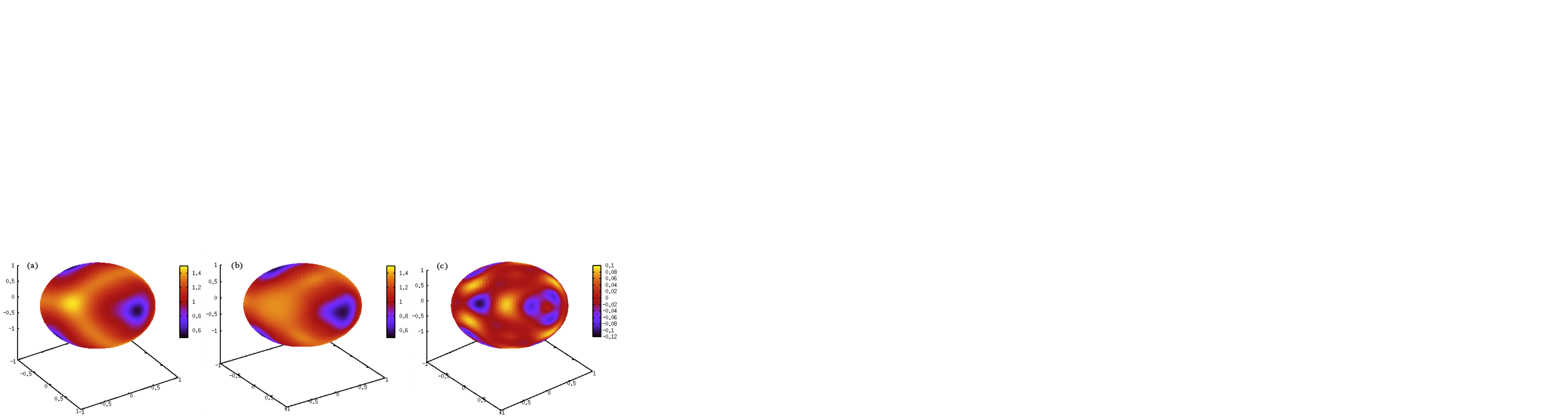}
 \includegraphics[width=15.8cm]{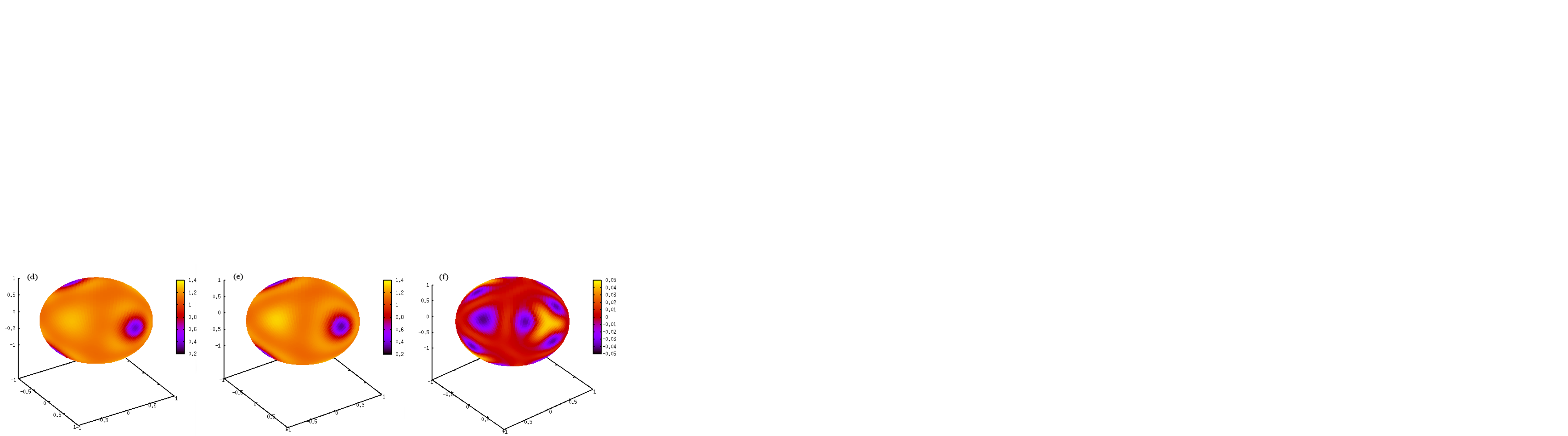}

\caption{(Color online) Particle density on the sphere surface for the two
states of the MR doublet: (a),(b) in the MR limit and (d),(e) in the
Coulomb case; (c) and (f) show the density difference between the
two states ($N=14$, $N_{\phi}=27$).}
\end{figure*}

In the tetrahedral case, right in the Coulomb limit, a higher-energy
state comes down in energy and nearly reaches the upper state of the
doublet. But Fig.~4(d) shows that this is actually not a problem:
The Coulomb spectrum is plotted as function of the strength of the
pinning potential and it is evident that the near degeneracy is easily
lifted by increasing the pinning strength. We also note that the two
states of the doublet are not degenerate in the presence of a Coulomb
interaction but we expect the splitting to vanish in the thermodynamic
limit.\cite{baraban2}

The curves denoted by ``Pfaffian'' in Figs.~4(a) and 4(b) show
the variational results for the splitting of the doublet, computed
using the states at $x=0.95$ (compare with Baraban \textit{et al.}
\cite{baraban2}): They overestimate the splitting in the Coulomb
limit, in the square geometry by a factor of $2.5$ and in the tetrahedral
geometry by $\sim 70 \%$.

Finally, Fig.~4(c) shows the evolution of the overlaps $\langle\psi_{i}(0)|\psi_{i}(x)\rangle$
of the two states $\psi_{i}(x)\,\,(i=0,1)$ of the doublet as the
interaction varies from Coulomb $V_{C}$ at $x=0$ to the vicinity
of the three-body limit $V_{3b}$ at $x=0.95$. We see that there
is no sign of an abrupt drop of the overlaps which could signal a
phase transition, and the values reached near the MR limit are reasonably
large for a system of this size ($N=14,N_{\phi}=27$).

All these results provide evidence in favor of adiabatic continuity
from the MR to the Coulomb limit for the two lowest-energy states
in systems with four localized QHs, suggesting that the non-Abelian
doublet, and the associated Majorana fermion, can be expected in the
limit of pure Coulomb interaction.

\subsection{Quasiholes localization with ``STM tip'' pinning}
\label{MRlimit}

In the AC investigations described so far, the pinning potential $V_{pin}(x)$
[in Eq.~(\ref{Vpin1})] was not able to localize QHs in the MR limit
(as discussed in section \ref{system}), but we approached this limit
near enough to recognize the lowest-lying states. One can pose the
question if the permanent presence of the Coulomb interaction (that
vanish as $1-x$ in the MR limit $x\to1$) could influence our results,
selecting exactly the ``right states'' from the zero energy states
at the MR point. To strenghten our evidences and to show a complete
AC between Coulomb and MR limit, we repeat here the calculations for
systems containing four QHs, using a different pinning potential
$V_{pin}(x)$ that allows to reach both limits: 
\begin{equation}
V_{pin}(x)=(1-x)V_{q}^{(1)}+xV_{q}^{(0)},\label{e/4}
\end{equation}
where $V_{q}$, interpolated between the $n=1$ and the $n=0$ LL
like the electron interaction, is the Coulomb potential of a pointlike
object, e.g., a STM tip, with a (negative) charge $q$, positioned
on the surface of the FQH sample.\cite{locpot} This potential repels
the electrons from its center and, if the charge $q$ is chosen in
an appropriate way, namely, in an interval near $-e/4$, it can only
localize $e/4$-charged QHs, thus avoiding the problem of the localization
of $e/2$-charged ``double QHs'' in the MR limit. This idea was
first used by Prodan and Haldane \cite{prodan09} in their investigation
of the (non-Abelian) braiding properties of MR QHs, although projecting
the pinning potential on the zero energy space of their Hamiltonian.

Figure 5 shows the results for the low-lying energy spectrum, as the
interaction is varied from the Coulomb ($x=0$) to the MR limit ($x=1$),
for systems with four QHs localized at the corners of a tetrahedron
by four such pinning potentials. The results in Fig.~5(a) are for
$N=12$ electrons and a localizing charge $q=-e/4$, while in Fig.~5(b)
for $N=14$ electrons \cite{maximum_size} and $q=-0.19e$: These
charges were chosen in order to obtain that the two states of the
MR doublet are exactly the lowest-lying states in the MR limit.

In both cases we see that (once $q$ is fixed) the two lowest energy
states remain the lowest over the whole parameter interval (for $N=12$
with a level crossing). The third energy level, corresponding to an
excitation in the pinning potential, is exactly threefold degenerate
(for symmetry reasons) and can thus be easily distinguished from the
two lowest states forming the MR doublet.

We note that this doublet has a finite energy splitting, not only
for pure Coulomb interaction, but also in the MR limit: This is caused
by the pinning potential (\ref{e/4}) that mixes in states with nonzero
energy. We also find that the energy separation between the upper
state of the doublet and the third (threefold degenerate) state can
be small, particularly in the MR limit, and depends on the charge
$q$ used in the pinning potential. For example for $N=14$ electrons,
using a localizing charge $q=-e/4$, the pinning potential is too
strong and at the MR point the threefold degenerate state is slightly
lower in energy than the upper state of the doublet; to obtain the
right order we chose a weaker potential, with $q=-0.19e$.

The AC between the Coulomb and the MR limit for the lowest energy
doublet shown in Fig.~5 thus supports the evidences of the previous
section in favor of a ``survival'' of the Majorana fermion till
the Coulomb point.

In Fig.~6 we also show the particle density on the surface of the
sphere for the two states of the MR doublet in the two limiting cases:
(a), (b) for the MR and (d), (e) for the Coulomb limit. The two states
of the doublet are indeed very similar, as also shown in (c) and (f),
where the density differences are plotted. Note that the density differences
in the MR case are approximately twice as large as the differences in the Coulomb
limit.

\section{Quasiholes braiding and fusion}

In this last section we show some preliminary results of QH
braiding and fusion in systems containing four localized QHs. We
perform this investigation for electron interactions near to the Coulomb
limit, but slightly increasing $v_{1}$, the Haldane pseudopotential
\cite{Haldane} that describes the interaction between two particles
having a relative angular momentum $1\hbar$. In our previous work
\cite{prl} we showed that, under such interaction modifications,
the system remains in the MR phase, even improving the overlap between
the exact GS and the Pfaffian wave function and simultaneously with
an enhancement of the gap. For the QH localization we use the pinning
potential introduced in the previous section [Eq.~(\ref{e/4})], setting
$x=0$, that is evaluating it fully in the $n=1$ LL, and chosing
a suitable value for the localizing charge $q$, as explained below.

Investigating QH braiding in systems with four QHs, the non-Abelian
nature of the MR doublet states should become manifest: Interchanging
the positions of two QHs by stepwise changing the location of their
pinning potentials (and keeping the other two QHs fixed), the two
states of the MR doublet should transform into each other, in particular,
switching their position in the energy spectrum. For this to happen,
during the QHs interchange process, an odd number of level crossings
between them is needed. In the following calculations, we engineer
an exact energy degeneracy for the two states of the doublet at the
midpoint of the QHs interchange process, by choosing suited values
for the localizing charge $q$ and the first Haldane pseudotential
$v_{1}$. Then, performing the interchange, we investigate if this
is indeed a crossing point and whether it is the only one or if others
arise.

\begin{figure}[t!]
 \includegraphics[width=1\columnwidth]{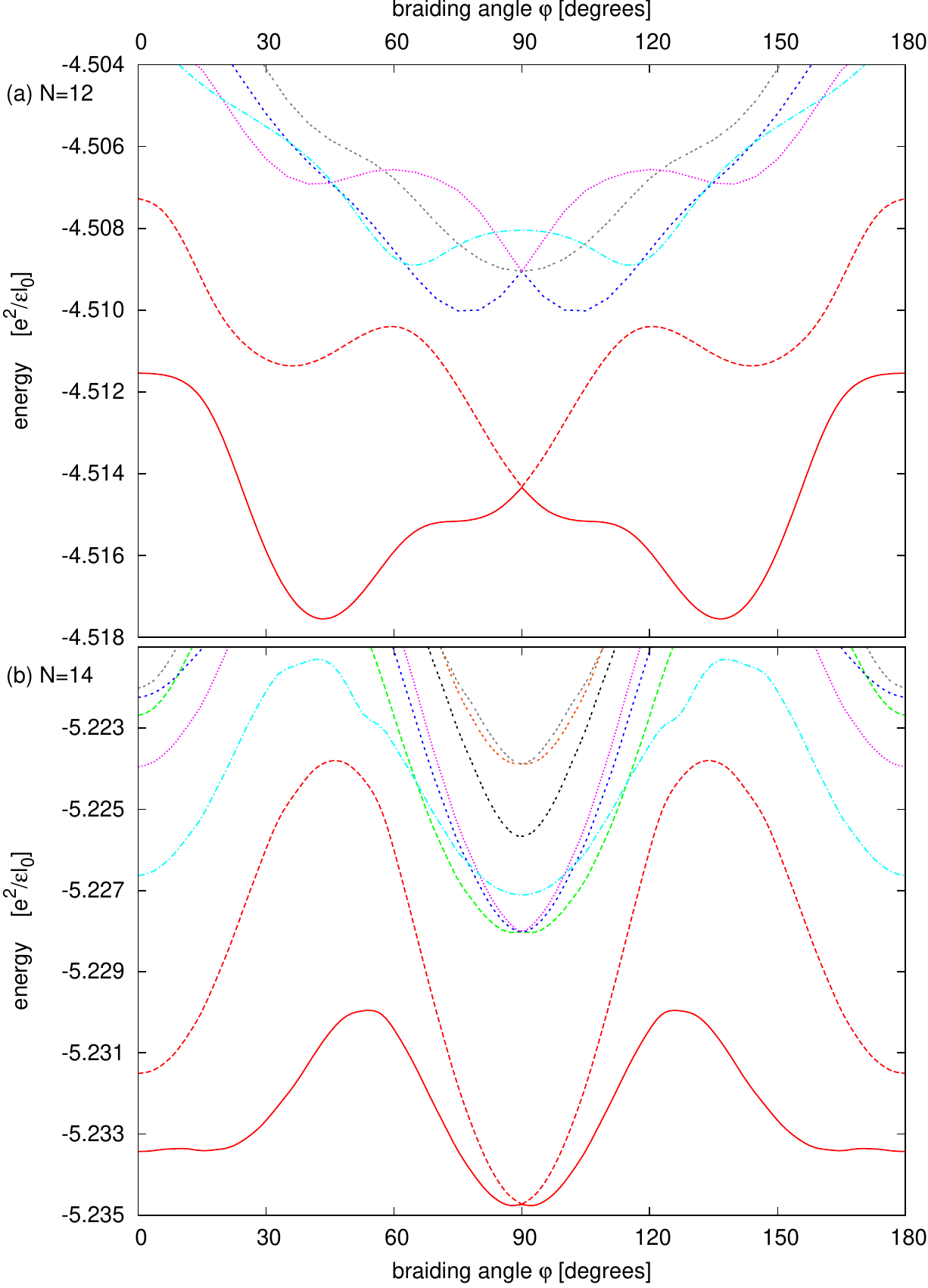}

\caption{(Color online) Braiding of two QHs (in the presence of two other
fixed QHs): low-lying energy spectrum as function of the braiding
angle $\varphi$, defined through Eq.~(\ref{brang}). The electron
interaction is near to the Coulomb limit (with a slightly increased
Haldane pseudopotential $v_{1}$) and the QHs are localized by
the pinning potential of Eq.~(\ref{e/4}), setting $x=0$. (a) $N=12$,
$q=-0.1715e$, $v_{1}=1.05v_{1}^{Coul}$, (b) $N=14$, $q=-0.19e$,
$v_{1}=1.075v_{1}^{Coul}$, where $v_{1}^{Coul}$ is the Coulomb value
of the first Haldane pseudopotential. }
\end{figure}

We chose as starting (and ending) configuration four QHs at the corners
of a rectangle: Two of them on the upper half-sphere, at the same
polar angle $\theta_{1,2}\equiv\theta_1=\theta_2=54.736^{\circ}$ and at the opposite azimuthal
angles $\varphi_{1}=90^{\circ},\varphi_{2}=270^{\circ}$; the other
two on the lower half-sphere at the polar angle 
$\theta_{3,4}\equiv\theta_3=\theta_4=125.264^{\circ}$
and azimuthal angles $\varphi_{3}=90^{\circ}$ and $\varphi_{4}=270^{\circ}$.
We keep QHs $3$ and $4$ fixed at their locations and, by stepwise
changing the positions of the pinning potentials, we rotate both QH
$1$ and $2$ around the vertical axis through the poles, 
\begin{equation}
\varphi_{1}(\varphi)=90^{\circ}+\varphi,\,\varphi_{2}(\varphi)=(270^{\circ}+\varphi)\bmod{360^{\circ}},\label{brang}
\end{equation}
keeping the polar angle $\theta_{1,2}=54.736^{\circ}$ constant,
until they exchange their original positions: $0^{\circ}\le\varphi\le180^{\circ}$.
This is a $180^{\circ}$-periodic process; we call $\varphi$ the
``braiding angle''.

The polar angles $\theta_{1,2}$ and $\theta_{3,4}$ are chosen in
such a way that at the midpoint of the rotation process, that is at
braiding angle $\varphi=90^{\circ}$, the four QHs are in a tetrahedral
configuration. The localizing charge $q$ and Haldane pseudopotential
$v_{1}$ (for the whole braiding process) were previously chosen,
such that at this point the two states of the MR doublet are degenerate
in energy: $q=-0.1715e$, $v_{1}=1.05v_{1}^{Coul}$ for $N=12$ and
$q=-0.19e$, $v_{1}=1.075v_{1}^{Coul}$ for $N=14$ electrons, where
$v_{1}^{Coul}$ is the Coulomb value of the pseudopotential $v_{1}$.
These are ``reasonable'' values for $q$ and $v_{1}$: The localized
QHs have charge $e/4$ and the electron interaction is modified such
that the system is still in the MR phase.

Figure 7 shows the evolution of the low lying energy spectrum as function
of the braiding angle $\varphi$ for (a) $N=12$ and (b) $N=14$ electrons.
In both cases we observe that the two states of the doublet remain
the lowest-energy states during the whole braiding process, never
mixing with higher-lying states or with each other (except at the
degeneracy point). Indeed the overlaps between the corresponding states
of the doublet for successive braiding steps are consistently high:
In the range $0.94$-$0.97$ for $N=12$ ($5^{\circ}$ steps) and
$0.97$-$0.99$ for $N=14$ ($3^{\circ}$ steps).

From the plots it is evident that the degeneracy point at $\varphi=90^{\circ}$
is indeed also a crossing point for the two doublet states: If one
follows the lines with a continuous derivative, at the degeneracy
point one goes from the lower to the upper state of the doublet (and
vice versa). No other crossing or mixing point arises and thus the
QH braiding causes an interchange in the MR doublet states: If one
starts at $\varphi=0^{\circ}$ in the lower-energy state, one ends
at $\varphi=180^{\circ}$ in the upper-energy state, as expected for
non-Abelian braiding statistics. Two such braidings are needed to
come back to the initial state.

We note that we can get an exact degeneracy of the MR doublet only
when the four QHs are in the tetrahedral configuration. Slightly
modifying the geometry of the configuration, a small gap opens; however,
if the speed of the braiding is sufficiently high, the states of the
MR doublet would still interchange at this point. This situation is
analogous to that studied by Thouless and Gefen,\cite{thoulessgefen}
concerning the crossing between the lowest-lying states as function
of the magnetic flux, for the quantum Hall effect at fractional fillings,
showing that these crossings are essential in order to get a fractional
charge. On the other hand if the QHs are moved in a strictly adiabatic
way, the system always remains in the lowest-energy state, without
crossing. We also wish to emphasize that the doublet degeneracy is
very different from the threefold degeneracy of the first excited
state: The former is obtained only by fine tuning of the parameters
$q$ and $v_{1}$, the latter is a purely geometric degeneracy, resulting
from the symmetric QH configuration and independent from interaction
and localization parameters.

The fact that the MR doublet states are not degenerate (except by
tuning at $\varphi=90^{\circ}$) in our small system diagonalizations
helps us to follow their evolution during the braiding (through the
evolution of the energy levels) and thus to recognize the crossing
point which leads to the interchange of states. However we expect
that in the thermodynamic limit the two states become exactly degenerate
during the whole braiding process, when the QHs are sufficiently
far apart from each other:
On one side this would make the fine tuning of the parameters $q$
and $v_1$ unnecessary, but on the other side it would make it impossible,
using our present method, to follow the evolution of the states under braiding
(however, it is still possible
with a study of the monodromy matrix, as was done by Prodan and Haldane
\cite{prodan09}); nevertheless, the interchange of the MR doublet
states should survive also in this limit.

Unfortunately we can get results for QH braiding only for small
system sizes \cite{maximum_size} and from them we cannot extract
useful information about the dependence of the energy splitting when
approaching the thermodynamic limit. The average splitting of the
MR doublet for $N=14$ is slightly smaller than for $N=12$, but this
is mainly caused by the different shape of the energy oscillations
as function of the braiding angle $\varphi$ in the two cases: For
$N=14$ the oscillations of the MR doublet states are almost ``in
phase'', while for $N=12$ they are ``out of phase'', thus giving
a larger average splitting.

\begin{figure}[t!]
 \includegraphics[width=0.96\columnwidth]{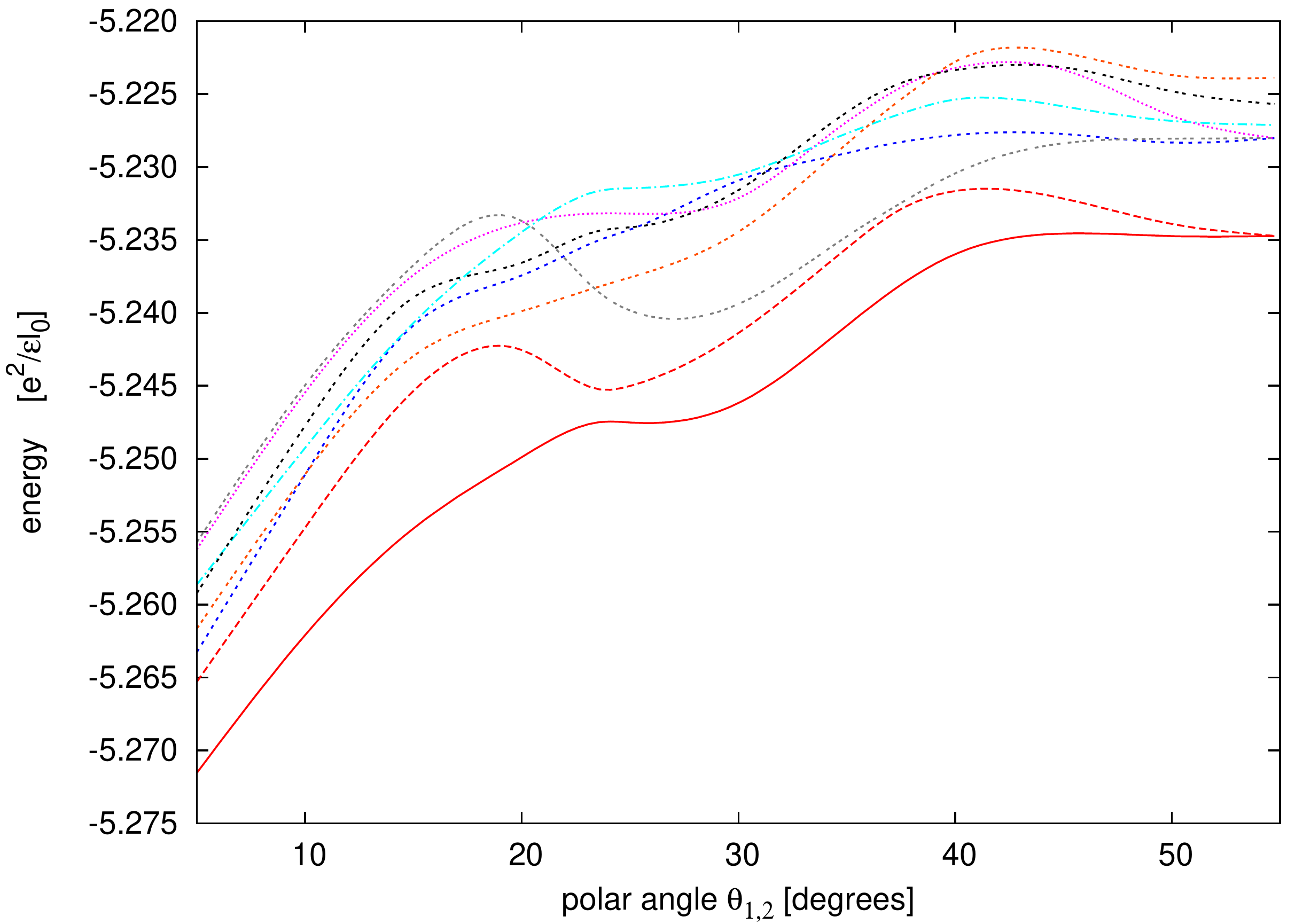}

\caption{(Color online) Fusion of two QHs at the north pole (in the presence
of two other fixed QHs): Low-lying energy spectrum for $N=14$ electrons
as function of $\theta_{1,2}\equiv\theta_1=\theta_2$, the polar angle of both the moving
QH $1$ and $2$. The electron interaction is near to the Coulomb
limit (with a slightly increased Haldane pseudopotential $v_{1}=1.075v_{1}^{Coul}$)
and the QHs are localized by the pinning potential of Eq.~(\ref{e/4}),
setting $x=0$ and $q=-0.19e$ ($v_{1}^{Coul}$ is the Coulomb value
of the first Haldane pseudopotential).}
\end{figure}

Finally we investigate the fusion of two QHs by bringing the pinning
potentials that localize them close to each other, in the presence
of two others fixed QHs. We start from the tetrahedral configuration
described above and we let fuse the two QHs in the upper half-sphere,
by taking them to the north pole, that is, by shrinking the polar angle
$\theta_{1,2}\equiv\theta_1=\theta_2$ from $54.736^{\circ}$ to $5^{\circ}$ at fixed azimuthal
angles. Figure 8 shows the low-lying energy spectrum as function of
the polar angle $\theta_{1,2}$ during this process, for $N=14$ electrons.
We see that, after that the degeneracy is lifted, the evolution of
the two lowest states is similar, with a relatively constant splitting
between them. This splitting (or at least a part of it) could come
from the intrinsic splitting of the MR doublet as two QHs are taken
together (its order of magnitude is the same as that obtained by Baraban
\textit{et al.}\cite{baraban2}). However, at approximately $\theta_{1,2}=37^{\circ}$,
the upper state of the MR doublet comes very near to the next higher-lying
state and strongly mixes with it \cite{mix} (but without crossing),
possibly losing a part of its character. Thus it is not fully clear
what happens at the MR doublet, and further investigations are needed.

\section{Conclusion}

Our results provide evidence in favor of adiabatic continuity from
the Moore-Read to the Coulomb limit for the two lowest-energy states
in systems with four localized QHs and thus the non-Abelian doublet
and associated Majorana fermion can be expected in the limit of pure
Coulomb interaction. Forces breaking particle-hole symmetry like $V_{3b}$
are actually present due to LL mixing \cite{bishara09,rezayi09} and
will favor either the Pfaffian \cite{MooreRead91} or the Antipfaffian.
\cite{levin07,lee07}

\begin{acknowledgments}
We acknowledge fruitful discussions with Nicholas d'Ambrumenil, J\"urg
Fr\"ohlich, Duncan Haldane, and Sankar Das Sarma, as well as the support
by the Swiss National Science Foundation and by the Institute for
Theoretical Physics at ETH, Zurich.
\end{acknowledgments}


\begin{thebibliography}{References}
\bibitem{maximum_size}$N=14$ is the maximum our resources allow:
Systems with $N_{QH}>2$ localized QHs break rotation symmetry. At
$N=14$ we have a Hilbert space dimension $N_{H}\approx10^{7}$ and
a number of couplings $V_{3b}$: $N_{C}\approx4\times10^{11}$. At $N=16$,
$N_{H}\approx6\times10^{8}$ and $N_{C}\approx3\times10^{12}$ which
is too large.
\bibitem{mix} At this point the ``order preserving'' overlaps
are of the order $0.76$, while the ``crossing'' overlaps of the
order $0.59$, for a $3^{\circ}$ step. (Typical overlaps between
corresponding states for successive steps are instead well above $0.9$.)
\end{thebibliography}
\end{document}